\documentclass[
  ,final            
  ,numberedheadings 
  ]
  {aipproc}

\layoutstyle{6x9}

\begin{document}

\title{The (n,$\gamma$) cross sections of short-living $s$-process
  branching points}

\author{K.\ Sonnabend}{
  address={Institut f\"ur Kernphysik, Technische Universit\"at
    Darmstadt, Germany}
}
\author{A.\ Mengoni}{
  address={CERN, Geneva, Switzerland}
}
\author{P.\ Mohr}{
  address={Institut f\"ur Kernphysik, Technische Universit\"at
    Darmstadt, Germany}
}
\author{T.\ Rauscher}{
  address={Institut f\"ur Physik, Universit\"at Basel, Switzerland}
}
\author{K.\ Vogt}{
  address={Institut f\"ur Kernphysik, Technische Universit\"at
    Darmstadt, Germany}
}
\author{A.\ Zilges}{
  address={Institut f\"ur Kernphysik, Technische Universit\"at
    Darmstadt, Germany}
}

\begin{abstract}
  An experimental method to determine the (n,$\gamma$) cross section of
  short-living $s$-process branching points using data of the inverse
  ($\gamma$,n) reaction is presented. The method was used to observe
  the branching point nucleus $^{95}$Zr because the elemental abundance
  patterns corresponding to this branching point cannot be reproduced
  by full stellar models and a possible error source is the neutron capture
  cross section of $^{95}$Zr. The analysis of the experiment is still
  under progress, we will outline the current status in this manuscript.
\end{abstract}

\maketitle


\section{Introduction}
\label{sec:intro}
The nucleosynthesis of the elements heavier than iron is today mainly
explained by three processes: $s$-, $r$-, and $p$-process. $s$- and
$r$-process are based on neutron-capture reactions with adjacent
$\beta$-decays while the $p$-process is governed by photodisintegration
reactions such as ($\gamma$,n), ($\gamma$,p), or ($\gamma$,$\alpha$).

The distinction between $s$- and $r$-process is due to the different
neutron densities $n_{\rm n}$ involved. During $s$-process
nucleosynthesis $n_{\rm n}$ is in the order of $10^8\ {\rm
  neutrons}/{\rm cm}^3$ while typical $r$-process sites deal with
$n_{\rm n} > 10^{20}\ {\rm neutrons}/{\rm cm}^3$ \cite{beer00}. Hence,
the neutron capture rates $\lambda_{({\rm n},\gamma)}$ are larger than
typical $\beta$-decay rates $\lambda_\beta$ during $r$-process
nucleosynthesis ($\lambda_{({\rm n},\gamma)} \gg \lambda_\beta$, $r$:
rapid). In the $s$-process the situation is the other way round
($\lambda_{({\rm n},\gamma)} \ll \lambda_\beta$, $s$: slow) and the
involved nuclei are close to the valley of $\beta$-stability.

However even in a $s$-prozess scenario, if the half-life $T_{1/2}$ is
long enough and the neutron capture cross section high enough another
neutron capture might take place and the path ``branches''
out. Therefore, these nuclides are called branching points of the
$s$-process. The branching ratio -- i.e. how often each of the paths
is taken -- determines the corresponding elemental abundance
patterns. Using a model for $s$-process nucleosynthesis and knowing
precisely the nuclear physics input (half-life $T_{1/2}$ and
Maxwellian averaged capture cross section (MACS) at typical
$s$-process temperatures $kT = 30\ {\rm keV}$) it is possible to
determine the astrophysical parameters temperature $T$ and neutron
density $n_{\rm n}$.

In the so-called classical approach temperature $T$ and neutron
density $n_{\rm n}$ are considered to be constant. Thus, three
different components produced with different astrophysical parameters
are needed to reproduce the observed abundances of $s$-only nuclei:
the weak component corresponds to mass numbers $A < 90$ while the
strong component only describes the termination of the $s$-process
path at lead and bismuth. In between these borders the isotopes belong
to the so-called main $s$-process \cite{kaep99}. The abundances of
$s$-only isotopes that are not affected by a branching can be
reproduced with a mean square deviation of about 3\% using this simple
model \cite{kaep90}. 

A more realistic approach is a full stellar model, e.g. the AGB star
model described in Ref.~\cite{luga03a}. These models need very precise
nuclear physics input data to reproduce the $s$-only
abundances. Especially the MACS of several branching points hamper the
reliability of the predictions: due to the lack of experimental data
the theoretical predictions of the MACS sometimes show a broad spread
(see \cite{bao00}).

A direct measurement of the MACS of the branching points is only
possible if their half-lives are in the order of years
(e.g. $^{147}$Pm with $T_{1/2} = 2.62\ {\rm yr}$ \cite{reif03}). In
the case of short-living branching points with half-lives of about a
dozen days or even less, adequate samples are not available. Thus,
direct measurements are possible.

To solve this problem we have investigated an experimental method
using the data of the inverse ($\gamma$,n) reaction as described in
Section~\ref{sec:expmeth}. Section~\ref{sec:prop} presents 
$^{95}$Zr focusing on astrophysical aspects as well as on the
constraints of our experimental method for this nuclide. The
present status of the analysis is summarized in
Section~\ref{subsec:status}. In Section~\ref{sec:summ}
we describe the next step in the data evaluation.

\section{Experimental method}
\label{sec:expmeth}
The aim of our experimental method is to constrain the theoretical
predictions of the MACS of short-living $s$-process branching
nuclei. Short-living under $s$-process conditions denotes
half-lives of about a dozen days or less, i.e.\ direct measurements are
not possible due to the lack of adequate samples. Therefore, our
method is based on data of the inverse ($\gamma$,n) reaction.

The first step is to measure the so-called energy-integrated cross
section $I_\sigma$ using the photoactivation technique:
\begin{equation}
\label{equ:int}
I_\sigma = \int^{E_{\rm max}}_{S_{\rm n}} \sigma(E)
\cdot N_\gamma(E,E_{\rm max}) \cdot {\rm d}E
\end{equation}
with $\sigma(E)$ being the ($\gamma$,n) cross section and
$N_\gamma(E,E_{\rm max})$ the experimental photon flux.

We use the monoenergetic electron beam provided by the S--DALINAC
\cite{rich00} to produce a continuous bremsstrahlung spectrum by fully
stopping the beam in a thick rotating copper target. The spectrum
ranges from $E=0$ to $E_{\rm e^-}$, so that $E_{\rm max}$ of
Eq.~\eqref{equ:int} equals $E_{\rm e}$. Due to the location of our
setup directly behind the injector of the S--DALINAC, our maximum
energy is limited to about $E_{\rm e^-} = 10\ {\rm MeV}$ presently
\cite{mohr99a}. While the shape of $N_\gamma(E,E_{\rm max})$ is well
known the absolute value has to be determined by a relative
measurement with gold as a calibration standard.

After irradiating the targets the yield $Y$ of the activation is
measured by counting the $\gamma$-rays emitted in the decay of the
produced radioactive nuclei with a HPGe-detector that is well-shielded
against background by several layers of lead. The yield $Y$ is
proportional to the energy-integrated cross section $I_\sigma$ with
the constant of proportionality being explained in detail in
Ref.~\cite{vogt01b}.  

Because the ($\gamma$,n) cross section of $^{197}$Au is known to behave like
\begin{equation}
\label{equ:au}
\sigma_{\rm Au}(E) = 146.2\ {\rm mb}\cdot \left(\frac{E - S_{\rm
      n}}{S_{\rm n}}\right)^{0.545}
\end{equation}
near threshold \cite{vogt02}, a predicted energy dependence of the
target's cross section $\sigma_{\rm target}(E)$ can be normalized to
the observed ratio $I_\sigma^{\rm target}/I_\sigma^{\rm Au}$ by a
correction factor $f$ for each measured energy $E_{\rm max}$. In doing
so, either a parametrization or a theoretical calculation of
$\sigma_{\rm target}(E)$ can be used. If $f$ depends on 
$E_{\rm max}$ the shape of $\sigma_{\rm target}(E)$ is not accurately
described (compare Eq.~\ref{equ:int}) and $f$ can not be used for the
further procedure without modifications.

Although the photoactivation technique is very sensitive,
several limitations concerning the activation reaction
$^A$X($\gamma$,n)$^{A-1}$X and the following decay of $^{A-1}$X have
to be taken into account. To test the energy dependece of the cross
section $\sigma_{\rm target}(E)$ it is necessary to measure at
different energies $E_{\rm max}$. Because our present setup is limited
to 10\ MeV, isotopes with neutron thresholds $S_{\rm n} > 9\ {\rm
  MeV}$ cannot be checked.

If one wants to use naturally composed targets, the fraction of the
isotope of interest must not be too small and the activation of other
isotopes should not produce too much background radiation when the
decay of $^{A-1}$X is observed. To reach a good peak-to-background
ratio in the measured spectra only a few $\gamma$-rays with high
branchings should be emitted in the decay of $^{A-1}$X. Furthermore,
the analyzed $\gamma$-decays should not be part of a multi-level decay
and their energies $E_\gamma$ should not allow self-absorption in the
target to avoid additional uncertainties in the analysis.

If the normalization of a theoretical prediction of $\sigma_{\rm
  target}(E)$ yields a constant factor $f$, we use $f$ to correct the
MACS predicted in the same model with equal parameters. This second
step of using the correction factor for the inverse reaction is based
on the detailed balance assumption. In our case, scaling the MACS with
a constant factor $f$ equals scaling the $\gamma$-ray strength
function. For further discussion see~\cite{sonn03a}.  

Although the corrected MACS is still based on a theoretical prediction,
its value is constrained by experimental knowledge of the inverse
reaction due to the determination of $f$. Thus, it is more reliable
than a value purely based on theory.

\section{The branching point $^{95}$Zr}
\label{sec:prop}
\subsection{Position on the $s$-process path}
\label{subsec:path}
\begin{figure}[h]
\includegraphics[width=\textwidth-2cm]{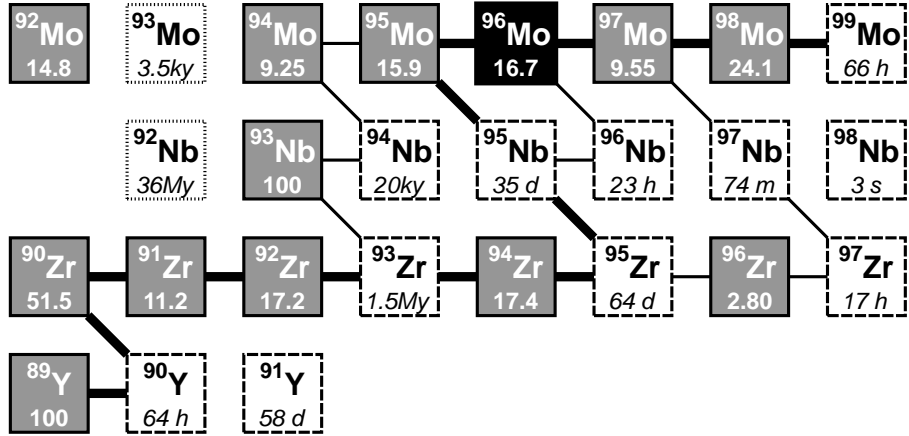}
\caption{The $s$-process path around the branching point
  $^{95}$Zr. The main flow of the $s$-process is marked with thick
  lines, thinner lines indicate branches. Grey shaded boxes illustrate
  stable isotopes with the numbers at the bottom of the boxes being
  their natural abundances. The black box of $^{96}$Mo stresses that
  we have an $s$-only nucleus. The white boxes display $\beta$-instable
  nuclei: dashed boundaries are used for $\beta^-$-decay, dotted lines
  are used to indicate $\beta^+$- or $\epsilon$-instable isotopes. In
  this cases the numbers at the bottom of the boxes are the half-lives
  of the ground-states.}
\label{fig:path}
\end{figure}

The branching point $^{95}$Zr is located near the borderline between
the weak and the main component of the $s$-process
\cite{kaep99}. Therefore, the prediction of the elemental abundance
patterns corresponding to this branching is a crucial test for the
validity of a full stellar model described in \cite{luga03a}.

Several problems concering the prediction of the zirconium abundance
patterns measured in SiC grains are reported in Ref.~\cite{luga03b}. The
uncertainty in the predicted MACS of $^{95}$Zr is mentioned as one of
the possible error sources due to the wide range of predicted values
from 23\ mb to 126\ mb at $kT = 30\ {\rm keV}$ (see~\cite{holm76},
\cite{touk90}, \cite{raus00}, and \cite{goriely}). Another reason for 
the deviation between the measured and predicted abundance patterns
might be the uncertainty in the half-life $T_{1/2}$ of $^{95}$Zr at
$s$-process temperatures. However, as confirmed by a recent
measurement \cite{sonn03d} the first excited level is at  
$E = 954\ {\rm keV}$ and thus, is not significantly populated at $kT =
30\ {\rm keV}$. Hence, the half-life of $^{95}$Zr does not depend on
temperature under $s$-process conditions and can be omitted as an
error source because of its small error: $T_{1/2}(^{95}{\rm Zr}) =
(64.032 \pm 0.006)\ {\rm d}$ \cite{ensdf}. 

\subsection{Constraints of the experiment}
\label{subsec:constr}
To use our experimental method the constraints described in
Section~\ref{sec:expmeth} have to be fullfilled for the activation
reaction $^{96}$Zr($\gamma$,n)$^{95}$Zr. The threshold of this
reaction is $S_{\rm n} = 7854\ {\rm keV}$, which can be easily reached
at our setup providing energies up to 10\ MeV. The abundance of
$^{96}$Zr in a naturally composed target is 2.8\% and thus, high
enough for the activation method.

\begin{figure}[h]
\includegraphics[width=\textwidth-2cm]{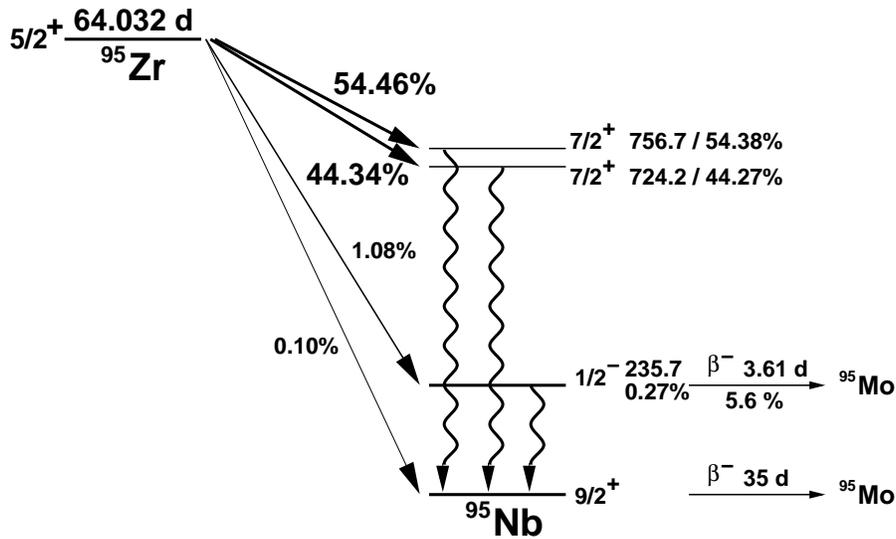}
\caption{Section of the decay scheme of $^{95}$Zr based on
  \cite{ensdf}}
\label{fig:decay}
\end{figure}

The properties of the decay of $^{95}$Zr are summarized in
Fig.~\ref{fig:decay}. The characteristics claimed in
Section~\ref{sec:expmeth} are achieved for $E_\gamma = 724.2\ {\rm
  keV}$ and $E_\gamma = 756.7\ {\rm keV}$.  

\begin{figure}[h]
\includegraphics[width=\textwidth-2cm]{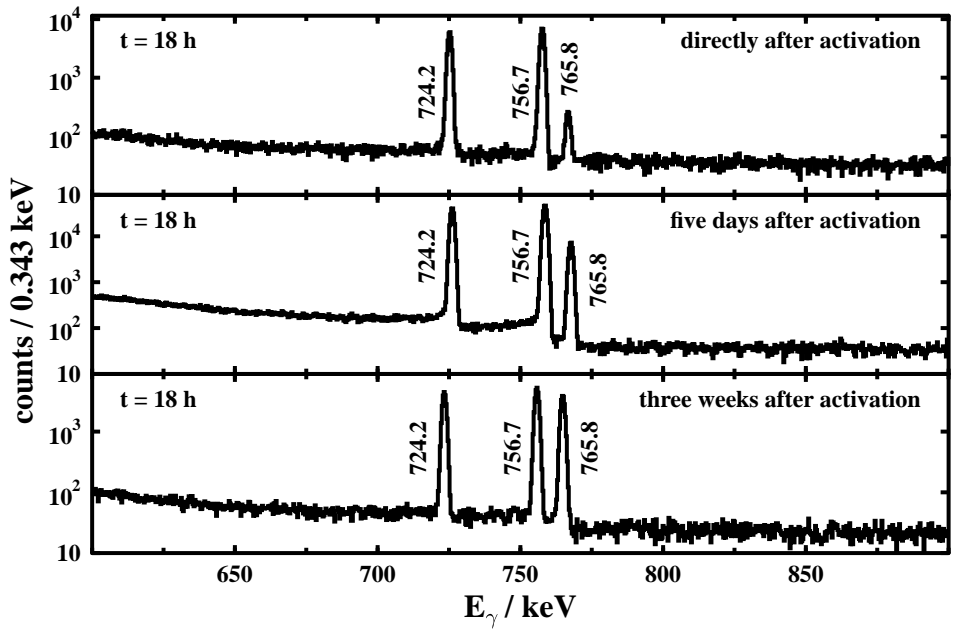}
\caption{Typical decay spectra of $^{95}$Zr each measured for 
  $t = 18\ {\rm h}$. From the top to the bottom panel the time between
  activation and start of the measurement is changing as
  indicated on the top right. For further explanation see text.
}
\label{fig:spectrum}
\end{figure}

Due to the half-life of $^{95}$Zr of about 64\ days it is mandatory to
observe the decay for several hours to reach good statistics. During
this time the $\beta^-$-decay of the daughter nucleus $^{95}$Nb also starts to
take place.  The assignment of the measured $\gamma$-rays to the
decays of $^{95}$Zr or $^{95}$Nb can either be done by a determination
of the observed half-life or by comparing their different behaviour in
time. As shown in Fig.~\ref{fig:spectrum} the ratio of the two peaks
at 724.2 and 756.7\ keV remains unchanged if the measurement takes
place different times after activation. Thus, both lines correspond to
the same decay and are identified to be the most prominent lines of
the decay of $^{95}$Zr. In contrast, the peak at 765.8\ keV becomes
more and more dominant with time. This behaviour is consistent with the one expected of a line
corresponding to the decay of the daughter nucleus
$^{95}$Nb. 
Actually, the energy $E_\gamma = 765.8\ {\rm keV}$ is
exactly the energy supposed for the most prominent line of this
decay. 

\subsection{Current status of analysis}
\label{subsec:status}
The photoactivation of zirconium has been done at six different
energies $E_{\rm max}$ ranging from 8325 to 9900\ keV. The lower limit
is due to the separation energy of the used calibration standard:
$S_{\rm n}({\rm Au}) = 8071\ {\rm keV}$. In a first approach, we used
a parametrization that describes the energy dependence of the
($\gamma$,n) cross section near threshold:
\begin{equation}
\label{equ:para}
\sigma(E) = \sigma_0 \cdot \left( \frac{E - S_{\rm n}}{S_{\rm n}}
\right)^p
\end{equation}
In this case, the exponent $p$ changes the shape of the energy
dependence, whereas the factor $\sigma_0$ is responsible for the
absolute value of $\sigma(E)$. Therefore, $\sigma_0$ corresponds to
the correction factor $f$ of Section~\ref{sec:expmeth}. Due to the
fact that this parametrization is derived from the time reversal
symmetry of (n,$\gamma$) and ($\gamma$,n) reactions, the exponent $p$
is connected to the wave-character of the emitted neutron: an exponent
of $p = 0.5 + l$ corresponds to an $l$-wave decay. Thus, exponents
less than 0.5 cannot be explained in this basic approach.

\begin{figure}[h]
\includegraphics[width=\textwidth-2cm]{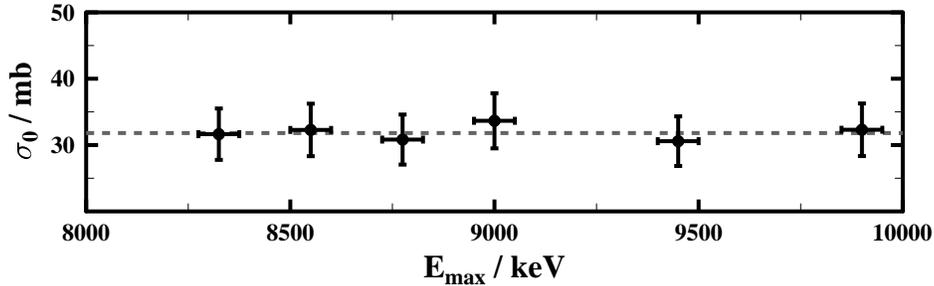}
\caption{Result of a least-mean-square-fit of Eq.~\eqref{equ:para} to
  the measured data at different energies $E_{\rm max}$. The dots are
  the results for $\sigma_0$ derived with an exponent of $p = 0.339$
  while the dashed line is the mean value $\sigma_0 = 31.81\ {\rm
  mb}$. The horizontal error bars are due to uncertainties in the beam
  energy.
}
\label{fig:fit}
\end{figure}

However, if a least-mean-square-fit of our data to this
parametrization is carried out using $\sigma_0$ and $p$ as free
parameters, we obtain $p = 0.339$ and $\sigma_0 = 31.81\ {\rm mb}$ as
shown in Fig.~\ref{fig:fit}. The single results $\sigma_0(E_{\rm max})$
fluctuate around the mean value, thus, the overall energy dependence of
$\sigma(E)$ seems to be described accurately.

The result $p < 0.5$ can be explained by assuming a resonance on top
of the smooth behaviour described by Eq.~\eqref{equ:para}. We measure
the energy integrated cross section $I_\sigma$ and compare it with the
result of Eq.~\eqref{equ:int} using $\sigma(E)$ of
Eq.~\eqref{equ:para}. Thus, we average the measured ($\gamma$,n) cross
section $\sigma(E)$ over the energy interval ranging from $S_{\rm n}$ to
$E_{\rm max}$. A resonance in this energy region would yield the 
measured bigger $I_\sigma$.

The smaller $p$, the steeper is the rise of the cross section at the
threshold energy $S_{\rm n}$. Hence, the calculated $I_\sigma$ becomes
bigger and suits the measured value. Thus, we accept the derived
values for $p$ and $\sigma_0$ as a description of the mean energy
dependence of the ($\gamma$,n) cross section: $\sigma(E) = 31.81\ {\rm
  mb} \cdot ((E - S_{\rm n}) / S_{\rm n})^{0.339}$.

\begin{figure}[h]
\includegraphics[width=\textwidth-2cm]{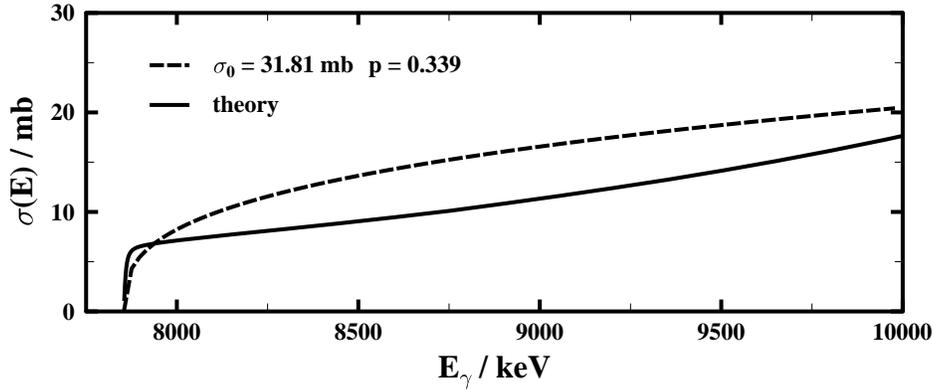}
\caption{Comparison of the energy dependence of the cross section
  $\sigma(E)$ for the reaction $^{96}$Zr($\gamma$,n). The solid line
  is a theoretical prediction based on Hauser-Feshbach
  calculations. The dashed line is the result of a
  least-mean-square-fit of the parametrization described in
  Eq.~\eqref{equ:para} to our data. 
}
\label{fig:sigma}
\end{figure}

Fig.~\ref{fig:sigma} shows the mean $\sigma(E)$ and a theoretical
prediction of the ($\gamma$,n) cross section $\sigma_{\rm theo}(E)$
based on Hauser-Feshbach calculations \cite{wolf51,haus52}.If the
shapes of the two energy dependences are compared, a significant
deviation is obvious: the theoretical prediction rises very steeply
close to the threshold, but its slope becomes flatter 50\ keV above
threshold. In contrast, our parametrization is not so steep near
threshold but more continuously rising.

Although the parametrization of $\sigma(E)$ describes only a mean
energy dependence of the cross section, the different behaviour
illustrates that the energy dependence of the theoretical prediction
does not suit the data as well. Obviously, no resonances are predicted
by theory. A correction factor $f$, derived as explained in
Section~\ref{sec:expmeth}, is only a mean correction. $f$ can compensate
the absence of resonances in theory. Furthermore, $f$ is a function of
$E_{\rm max}$. Therefore, the MACS of $^{95}$Zr predicted by this
theory cannot simply be corrected by this mean value of $f$ but the
(n,$\gamma$) cross section must be corrected by $f(E_\gamma)$ before
calculating the MACS. $f(E_\gamma)$ can be calculated weighting
$f(E_{\rm max})$ by $\sigma(E) \cdot N_\gamma(E,E_{\rm max})$. For
this purpose, it is necessary to measure the ($\gamma$,n) cross
section as close as possible to the reaction threshold to locate the
position of the resonances. 

So far, the experiment was limited by the threshold of the
calibration standard. The lowest measured energy was 
$E_{\rm max} = 8325\ {\rm keV}$ i.e.\ still about 500\ keV
above the threshold $S_{\rm n}$ of $^{96}$Zr. In order to measure as
close as possible near the target's threshold the neutron separation
energy of a new calibration standard has to be lower.

In addition to a low neutron separation energy the calibration standard
has to fullfill other characteristics: despite the general features
mentioned in Section~\ref{sec:expmeth}, the half-life $T_{1/2}$ of the
produced radioactive isotope has to be short so that the counting can
be done in short times with good statistics.

We decided to use $^{187}$Re with a neutron separation
energy of only $S_{\rm n} = 7359\ {\rm keV}$ and a natural abundance
of 62.6\%. The half-life of the produced $^{186}$Re is $T_{1/2} =
3.72\ {\rm d}$ and allows short measuring times although the branching
of the two dominant $\gamma$-decays is quite low: $E_\gamma = 137.2\
{\rm keV}$ with 9.42\% and  $E_\gamma = 122.6\ {\rm keV}$ with
0.6\%. Self-absorption in the rhenium target limits the amount of
material, however, based on experience of former experiments
\cite{sonn03a} very thin rhenium targets with a thickness of $d = 50\
\mu{\rm m}$ should provide enough activity after a typical activation. 

\begin{figure}[h]
\includegraphics[width=\textwidth-2cm]{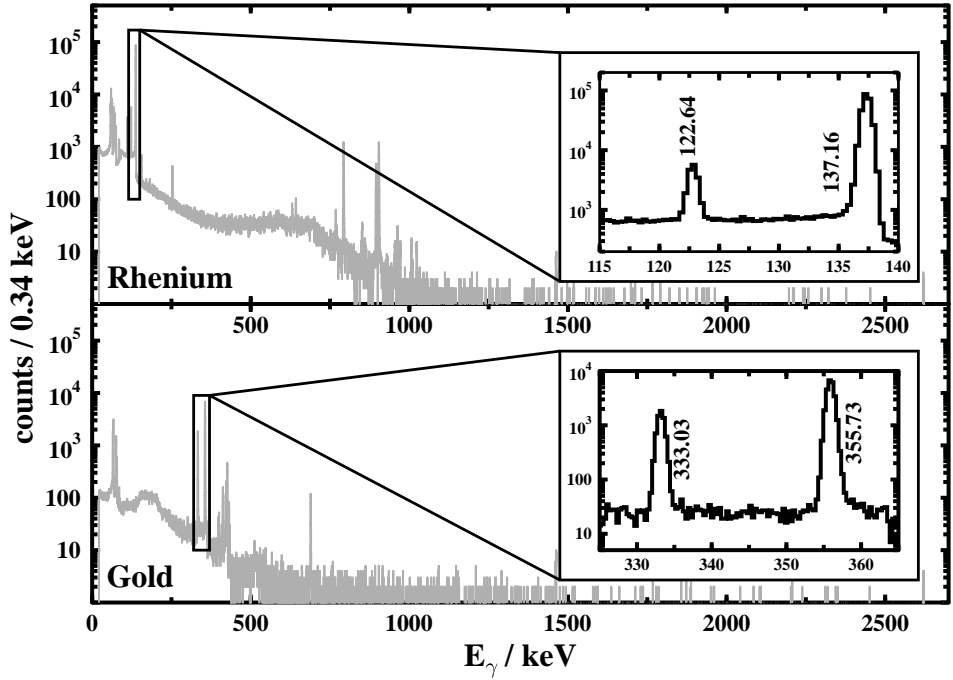}
\caption{Comparison of the decay spectra of the calibration standards
  gold and rhenium. The inlays show the $\gamma$ peaks that are used
  for the calibration. The peak-to-background ratio is in the same
  order of magnitude for both standards.
}
\label{fig:cali}
\end{figure}

$^{187}$Re was measured in May 2003 and the analysis is still
under progress. Fig.~\ref{fig:cali} compares the spectra of rhenium
and gold measured after an activation at $E_{\rm max} = 8325\ {\rm
  keV}$. The upper panel shows the spectrum of rhenium measured for
three hours directly after the activation. The peaks between 750 and
1000\ keV correspond to the decay of $^{184}$Re that is produced by
$^{185}$Re($\gamma$,n). Due to these lines the two peaks of the decay
of $^{186}$Re at $E_\gamma = 137.2\ {\rm keV}$ and 
$E_\gamma = 122.6\ {\rm keV}$ are located above a high
background. However, as shown in the inlay the peak-to-background
ratio of both lines is good enough to analyze the peak volumina with a
low statistical error. For comparison the inlay in the lower panel
shows the two most prominent lines of a gold spectrum also measured
for three hours. The peak-to-background ratio is in the same order of
magnitude. 

\section{Summary and Outlook}
\label{sec:summ}
The precise knowledge of the (n,$\gamma$) cross section of the
$s$-process branching point $^{95}$Zr is a crucial input parameter to
test the reliability of the AGB star model. Due to the very wide range of
the theoretically predicted values a measurement observing the inverse
reaction with the photoactivation technique was carried out at the
S--DALINAC in June 2002.

During the analysis of the data several problems occured:
The energy dependence of the ($\gamma$,n) cross section of $^{96}$Zr
extracted from the data differs significantly from the theoretical
prediction. In our opinion, the deviation is due to resonances that
are not included in the used theory. Therefore, the experimental
method described in Section~\ref{sec:expmeth} determines an energy
dependent correction factor $f(E_\gamma)$ that cannot be used to
correct the predicted MACS of $^{95}$Zr without further studies.

Hence, the experiment was extended in May 2003 by measuring the
($\gamma$,n) cross section of $^{96}$Zr closer to the reaction
threshold. Therefore, a new calibration standard with lower neutron
separation energy was measured during the same beam-time. The analysis
of the new standard $^{187}$Re is still under progress, preliminary
results like the spectrum shown in Fig.~\ref{fig:cali} are encouraging
that it will work as well as our first standard $^{197}$Au.

Once a description of the ($\gamma$,n) cross section of $^{187}$Re
near threshold is available the new data point of $^{96}$Zr taken at
$E_{\rm max} = 8100\ {\rm keV}$ has to be added to the existing
analysis. Therewith, the discrepancy between the experimental result
and the theoretical prediction will hopefully disappear or at least
become explainable. The outcome of the whole analysis that is the
experimentally confirmed MACS of $^{95}$Zr can then be used in the AGB
star model and might solve the problems in explaining the observed
zirconium abundance patterns in SiC grains.

In future, we will take more data points at different energies 
$E_{\rm max}$ in order to reduce difficulties due to resonances on top
of the cross section. If the distance between two data points $\Delta
E_{\rm max}$ is smaller, the determination of the location of a
resonance is possible with higher precision.


\begin{theacknowledgments}
  The authors would like to thank R.\ Gallino (Torino, Italy) and F.\
  K\"appeler (Karlsruhe, Germany) for fruitful discussions concerning
  $s$-process modelling. This work was supported by the Deutsche
  Forschungsgemeinschaft (contracts Zi 510/2-2 and SFB 634) and
  Swiss NSF (grants 2124-055832.98, 2000-061822.00, 2024-067428.01).
\end{theacknowledgments}


\bibliographystyle{aipproc}   

\bibliography{sonnabend}

\IfFileExists{\jobname.bbl}{}
 {\typeout{}
  \typeout{******************************************}
  \typeout{** Please run "bibtex \jobname" to optain}
  \typeout{** the bibliography and then re-run LaTeX}
  \typeout{** twice to fix the references!}
  \typeout{******************************************}
  \typeout{}
 }

\end{document}